\documentclass[a4paper]{jpconf}
\usepackage{graphicx}
                                                                                
\newcommand{\be}{\begin{equation}}
\newcommand{\ee}{\end{equation}}
\newcommand{\ba}{\begin{eqnarray}}
\newcommand{\ea}{\end{eqnarray}}
\newcommand{\ban}{\begin{eqnarray*}}
\newcommand{\ean}{\end{eqnarray*}}
                                                                                
\begin{document}

\title{Thermalization vs. Isotropization \\
\& Azimuthal Fluctuations\footnote{presented at the workshop 
{\it Correlations and Fluctuations in Relativistic Nuclear Collisions}, 
MIT, April 21-23, 2005}}

\author{Stanislaw Mr\'owczy\'nski}
                                                                                
\address{Institute of Physics, \'Swi\c etokrzyska Academy 
ul. \'Swi\c etokrzyska 15, PL - 25-406 Kielce, Poland 
and So\l tan Institute for Nuclear Studies 
ul. Ho\.za 69, PL - 00-681 Warsaw, Poland}
                                                                                
\ead{mrow@fuw.edu.pl}

\begin{abstract}

Hydrodynamic description requires a local thermodynamic equilibrium 
of the system under study but an approximate hydrodynamic behaviour is 
already manifested when a momentum distribution of liquid components 
is not of equilibrium form but merely isotropic. While the process of
equilibration is relatively slow, the parton system becomes isotropic 
rather fast due to the plasma instabilities. Azimuthal fluctuations 
observed in relativistic heavy-ion collisions are argued to distinguish 
between a fully equilibrated and only isotropic parton system produced 
in the collision early stage.

\end{abstract}

%%%%%%%%%%%%%%%%%%%%%%%%%%%%%%%%%%%%%%%%%%%%%%%%%%%%%%%%%%%%%%%%%%%%%%%%%%%
                                                                                
\section{Introduction}
                                                                                
%%%%%%%%%%%%%%%%%%%%%%%%%%%%%%%%%%%%%%%%%%%%%%%%%%%%%%%%%%%%%%%%%%%%%%%%%%%

A matter created in relativistic heavy-ion collisions manifests
a strongly collective hydrodynamic behaviour \cite{Heinz:2005ja},
particularly evident in studies of the so-called elliptic 
flow \cite{Alt:2003ab,Adler:2002pu,Adler:2003kt,Back:2004zg}.
Hydrodynamic description requires, strictly speaking, a local thermal 
equilibrium and experimental data on the particle spectra and the
elliptic flow suggest, when analysed within the hydrodynamic model, 
that an equilibration time of the parton\footnote{The term `parton' 
is used to denote a fermionic (quark) or bosonic (gluon) excitation 
of the quark-gluon plasma.} system produced at the collision early 
stage is as short as 0.6 ${\rm fm}/c$ \cite{Heinz:2004pj}. Such 
a fast equilibration can be explained assuming that the
quark-gluon plasma is strongly coupled \cite{Shuryak:2004kh}.
However, high-energy density in the collision early stage, when the
elliptic flow is generated \cite{Sorge:1998mk}, allows one to believe
that the plasma is then weakly coupled due to the asymptotic
freedom. Calculations, which assume that the parton-parton collisions 
are responsible for the equilibration of the weakly interacting plasma, 
provide an equilibration time of at least 2.6 ${\rm fm}/c$ 
\cite{Baier:2002bt}. To thermalize the system one needs either 
a few hard collisions of the momentum transfer of order of the 
characteristic parton momentum\footnote{Although I consider 
anisotropic systems, the characteristic momentum in all directions 
is assumed to be of the same order.}, which we denote here as $T$ 
(as the temperature of equilibrium system), or many collisions of 
smaller transfer. As discussed in {\it e.g.} \cite{Arnold:1998cy}, 
the inverse time scale of the collisional equilibration is of order 
$g^4 {\rm ln}(1/g)\,T$ where $g$ is the QCD coupling constant. 
However, it has been argued that the equilibration is speeded up 
by instabilities generated in an anisotropic quark-gluon plasma 
\cite{Mrowczynski:xv,Rebhan:2004ur,Arnold:2004ti}, as growth of the 
unstable modes is associated with the system's isotropization. The 
characteristic inverse time of instability development is roughly of 
order $gT$ for a sufficiently anisotropic momentum distribution
\cite{Mrowczynski:xv,Randrup:2003cw,Romatschke:2003ms,Arnold:2003rq}.
Thus, the instabilities are much `faster' than the hard collisions
in the weak coupling regime. Very recent classical simulation 
\cite{Dumitru:2005gp} indeed shows effectiveness of the instabilities
driven isotropization.
                                                                                
The isotropization should be clearly distinguished from the
equilibration process \cite{Mrowczynski:xv}. The instabilities driven
isotropization is a mean-field reversible phenomenon which is not
accompanied with entropy production. Therefore, the collisions,
which are responsible for the dissipation, are needed to reach the
equilibrium state of maximal entropy. The instabilities contribute
to the equilibration indirectly, reducing relative parton momenta
and increasing the collision rate.

It has been recently observed that the hydrodynamic collective behaviour
does not actually require local thermodynamic equilibrium but a merely
isotropic momentum distribution of liquid components \cite{Arnold:2004ti}.
Thus, there is a question whether a quark-gluon plasma, which is 
equilibrated nearly immediately after its production as advocated in
\cite{Shuryak:2004kh}, can be distinguished from the parton system
which slowly evolves towards equilibrium being isotropized fast. I argue 
here that measurements of azimuthal fluctuations, which are generated at 
the early stage of heavy-ion collisions, can help to distinguish the 
two scenarios.

In the first part of my talk I review the instabilities driven isotropization. 
I discuss how the unstable modes are initiated and what is a mechanism 
responsible for their growth. Dispersion relations of the unstable modes 
are considered, and it is explained why the development of instabilities
is associated with the system's isotropization. In the second part of my 
talk I discuss the azimuthal fluctuations, arguing that the fluctuations 
generated in the non-equilibrium isotropic system are much larger than
those in the fully equilibrated plasma. Two possible measurements are
proposed.

%%%%%%%%%%%%%%%%%%%%%%%%%%%%%%%%%%%%%%%%%%%%%%%%%%%%%%%%%%%%%%%%%%%%%%%%%%%
                                                                                
\section{Instabilities driven isotropization}
                                                                                
%%%%%%%%%%%%%%%%%%%%%%%%%%%%%%%%%%%%%%%%%%%%%%%%%%%%%%%%%%%%%%%%%%%%%%%%%%%

Temporal evolution of the electron-ion plasma is plagued by a large 
variety of instabilities. Those caused by coordinate space 
inhomogeneities, in particular by the system's boundaries, are usually 
called the hydrodynamic instabilities while those due to non-equilibrium 
momentum distribution of plasma particles the kinetic instabilities. 
Hardly anything is known about hydrodynamic instabilities of the 
quark-gluon plasma, and I will not speculate about their possible role 
in the system's dynamics. The kinetic instabilities are initiated either 
by the charge or current fluctuations. In the first case, the electric 
field (${\bf E}$) is longitudinal (${\bf E} \parallel {\bf k}$, where 
${\bf k}$ is the wave vector), while in the second case the field 
is transverse (${\bf E} \perp {\bf k}$). For this reason, the kinetic 
instabilities caused by the charge fluctuations are usually called 
{\it longitudinal} while those caused the current fluctuations 
{\it transverse}. Since the electric field plays a crucial role in 
the longitudinal mode generation, the longitudinal instabilities are 
also called {\it electric} while the transverse ones {\it magnetic}. 

In the non-relativistic plasma the electric instabilities are usually
much more important than the magnetic ones as the magnetic effects
are suppressed by the factor $v^2/c^2$ where $v$ is the particle's
velocity. In the relativistic plasma both types of similar strength. 
The electric instabilities occur when the momentum distribution has 
more than one maximum while a sufficient condition for the magnetic 
instabilities is, as discussed below, anisotropy of the momentum 
distribution. For this reason, the magnetic unstable mode, which 
is also called Weibel or filamentation instability \cite{Wei59}, 
was argued long ago to be relevant for equilibration of the 
quark-gluon plasma produced in relativistic heavy-ion collisions 
\cite{Mrowczynski:xv}. In the remaing part of the section, I am going 
to explain in detail why the filamentation is relevant and how it 
speeds up the process of plasma thermalization.

%%%%%%%%%%%%%%%%%%%%%%%%%%%%%%%%%%%%%%%%%%%%%%%%%%%%%%%%%%%%%%%%%%%%%%%%%%%
                                                                                
\subsection{Seeds of the filamentation}
                                                                                
%%%%%%%%%%%%%%%%%%%%%%%%%%%%%%%%%%%%%%%%%%%%%%%%%%%%%%%%%%%%%%%%%%%%%%%%%%%

Let me first discuss how the unstable transverse modes are initiated. 
For this purpose I consider a parton system which is homogeneous but 
the parton momentum distribution is, in general, not of the equilibrium 
form, it is {\em not} isotropic. The system is on average locally 
colourless but colour fluctuations are possible. Therefore, 
$\langle j^{\mu}_a (x)\rangle = 0$ where $j^{\mu}_a (x)$ is a local
colour four-current in the adjoint representation of ${\rm SU}(3)$ 
gauge group with $\mu=0,1,2,3$ and $a =1,2,3, \dots ,8$ being the Lorentz 
and colour index, respectively; $x=(t,{\bf x})$ denotes a four-position
in the coordinate space.  
 
As discussed in detail in \cite{Mrowczynski:1996vh}, the current 
correlator for a classical system of non-interacting quarks and 
gluons is 
\ba
\label{cur-cor-x}
M^{\mu \nu}_{ab} (t,{\bf x}) &\buildrel \rm def \over =& 
\langle j^{\mu}_a (t_1,{\bf x}_1) j^{\nu}_b (t_2,{\bf x}_2) \rangle 
= {1 \over 8} \,g^2\; \delta^{ab} 
\int {d^3p \over (2\pi )^3} \; {p^{\mu} p^{\nu} \over E_p^2} \;
f({\bf p}) \; \delta^{(3)} ({\bf x} -{\bf v} t)  \;,
\ea
where $(t,{\bf x}) \equiv (t_2-t_1,{\bf x}_2-{\bf x}_1)$ and the 
effective parton distribution function $f({\bf p})$ equals 
$n({\bf p}) + \bar n({\bf p}) + 6 n_g({\bf p})$ with $n({\bf p})$, 
$\bar n({\bf p})$ and $n_g({\bf p})$ giving the average colourless
distribution function of quarks $Q^{ij}(x,{\bf p})= \delta^{ij} n({\bf p})$, 
antiquarks $\bar Q^{ij}(x,{\bf p})= \delta^{ij} \bar n({\bf p})$, and 
gluons $G^{ab}(x,{\bf p})= \delta^{ab}n_g({\bf p})$. We note that
the distribution function of (anti-)quarks belongs to the fundamental
representation of the ${\rm SU}(3)$ gauge while that of gluons to
the adjoint representation. Therefore, $i,j=1,2,3$ and $a,b =1,2,...,8$.

Due to the average space-time homogeneity, the correlation tensor 
(\ref{cur-cor-x}) depends only on the difference 
$(t_2-t_1,{\bf x}_2-{\bf x}_1)$. The space-time points $(t_1,{\bf x}_1)$ 
and $(t_2,{\bf x}_2)$ are correlated in the system of non-interacting 
particles if a particle travels from $(t_1,{\bf x}_1)$ to $(t_2,{\bf x}_2)$. 
For this reason the delta  $\delta^{(3)} ({\bf x} - {\bf v} t)$ is 
present in the formula (\ref{cur-cor-x}). The momentum integral of the
distribution function simply represents the summation over particles.
The fluctuation spectrum is found as a Fourier transform of the tensor 
(\ref{cur-cor-x}) {\it i.e.} 
\be
\label{cur-cor-k}
M^{\mu \nu}_{ab} (\omega ,{\bf k}) = {1 \over 8} \,g^2\; \delta^{ab} 
\int {d^3p \over (2\pi )^3} \; 
{p^{\mu} p^{\nu} \over E_p^2} \; f({\bf p})  \;
2\pi \delta (\omega -{\bf kv}) \;.
\ee

To compute the fluctuation spectrum, the parton momentum distribution
has to be specified. Such calculations with two forms of the anisotropic 
momentum distribution are presented in \cite{Mrowczynski:1996vh}. Here 
I only qualitatively discuss Eqs.~(\ref{cur-cor-x},\ref{cur-cor-k}). 
I assume that the momentum distribution is elongated in, say, the $z$ 
direction. Then, Eqs.~(\ref{cur-cor-x},\ref{cur-cor-k}) clearly show that 
the corelator $M^{zz}$ is larger than  $M^{xx}$ or  $M^{yy}$. It also 
clear that $M^{zz}$ is the largest when the wave vector ${\bf k}$ is 
along the direction of the momentum deficit. Then, the delta function 
$\delta (\omega -{\bf kv})$ does not much constrain the integral in 
Eq.~(\ref{cur-cor-k}). Since the momentum distribution is elongated 
in the $z$ direction, the current fluctuations are the largest when 
the wave vector ${\bf k}$ is the $x\!-\!y$ plane. Thus, I conclude 
that some fluctuations in the anisotropic system are large, much larger
than in the isotropic one and that anisotropic system has a natural
tendency to split into the current filaments parallel to the direction 
of the momentum surplus. These currents are seeds of the filamentation
instability.

%%%%%%%%%%%%%%%%%%%%%%%%%%%%%%%%%%%%%%%%%%%%%%%%%%%%%%%%%%%%%%%%%%%%%%%%%%%

\subsection{Mechanism of filamentation}

%%%%%%%%%%%%%%%%%%%%%%%%%%%%%%%%%%%%%%%%%%%%%%%%%%%%%%%%%%%%%%%%%%%%%%%%%%%

Let me now explain in terms of elementary physics why the fluctuating 
currents, which flow in the direction of the momentum surplus, can grow
in time. To simplify the discussion, which follows \cite{Mrowczynski:1996vh}, 
I consider an electromagnetic anisotropic system. The form of the fluctuating 
current is chosen to be
\be
\label{flu-cur}
{\bf j}(x) = j \: \hat {\bf e}_z \: {\rm cos}(k_x x) \;,
\ee
where $\hat {\bf e}_z$ is the unit vector in the $z$ direction.
As seen in Eq.~(\ref{flu-cur}), there are current filaments of 
the thickness $\pi /\vert k_x\vert$ with the current flowing in the 
opposite directions in the neighbouring filaments. 

\begin{figure}
\begin{center}
\includegraphics[width=8cm]{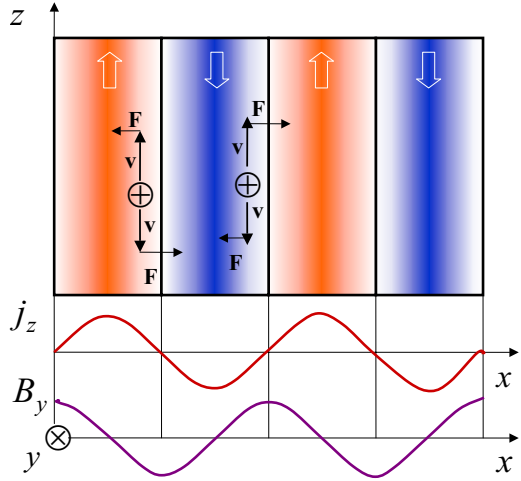}
\caption{The mechanism of filamentation instability.}
\end{center}
\end{figure}

The magnetic field generated by the current (\ref{flu-cur}) is given as
\ban
{\bf B}(x) = {j \over k_x} \: \hat {\bf e}_y \: {\rm sin}(k_x x) \;,
\ean
and the Lorentz force acting on the partons, which fly along the $z$
direction, equals
\ban
{\bf F}(x) = q \: {\bf v} \times {\bf B}(x) = 
- q \: v_z \: {j \over k_x} \: \hat {\bf e}_x \: {\rm sin}(k_x x) \;,
\ean
where $q$ is the electric charge. One observes, see Fig.~1, that the 
force distributes the partons in such a way that those, which positively 
contribute to the current in a given filament, are focused in the 
filament centre while those, which negatively contribute, are moved 
to the neighbouring one. Thus, the initial current is growing.

%%%%%%%%%%%%%%%%%%%%%%%%%%%%%%%%%%%%%%%%%%%%%%%%%%%%%%%%%%%%%%%%%%%%%%%%%%%

\subsection{Dispersion equation}

%%%%%%%%%%%%%%%%%%%%%%%%%%%%%%%%%%%%%%%%%%%%%%%%%%%%%%%%%%%%%%%%%%%%%%%%%%%

The Fourier transformed chromodynamic field $A^{\mu}(k)$ satisfies the 
equation of motion as
\be
\label{eq-A}
\Big[ k^2 g^{\mu \nu} -k^{\mu} k^{\nu} - \Pi^{\mu \nu}(k) \Big] 
A_{\nu}(k) = 0 \;,
\ee
where $\Pi^{\mu \nu}(k)$ is the polarization tensor or gluon self-energy 
which is discussed later on. A general plasmon dispersion equation is of 
the form 
\be
\label{dispersion-pi}
{\rm det}\Big[ k^2 g^{\mu \nu} -k^{\mu} k^{\nu} - \Pi^{\mu \nu}(k) \Big] 
 = 0 \;.
\ee
Equivalently, the dispersion relations are given by the positions of 
poles of the effective gluon propagator. Due to the transversality of 
$\Pi^{\mu \nu}(k)$ ($k_\mu \Pi^{\mu \nu}(k) = k_\nu \Pi^{\mu \nu}(k)=0$) 
not all components of $\Pi^{\mu \nu}(k)$ are independent from each other, 
and consequently the dispersion equation (\ref{dispersion-pi}), which 
involves a determinant of $4\times4$ matrix, can be simplified to the
determinant of $3\times3$ matrix. For this purpose I introduce the 
colour permittivity tensor $\epsilon^{lm}(k)$ where the indices 
$l,m,n = 1,2,3$ label three-vector and tensor components. Because of 
the relation 
$$
\epsilon^{lm}(k) E^l(k) E^m(k) = \Pi^{\mu \nu} (k) A_{\mu}(k) A_{\nu}(k)\;,
$$
where ${\bf E}$ is the chromoelectric vector, the permittivity can be 
expressed through the polarization tensor as
$$
\epsilon^{lm}(k) = \delta^{lm} + {1 \over \omega^2} \Pi^{lm}(k) \;.
$$
Then, the dispersion equation gets the form
\be
\label{dispersion-g}
{\rm det}\Big[ {\bf k}^2 \delta^{lm} -k^l  k^m 
- \omega^2 \epsilon^{lm}(k)  \Big]  = 0 \,.
\ee
The relationship between Eq.~(\ref{dispersion-pi}) and 
Eq.~(\ref{dispersion-g}) is most easily seen in the Coulomb 
gauge when $A^0 = 0$ and ${\bf k} \cdot {\bf A}(k)=0$. Then, 
${\bf E} = i\omega {\bf A}$ and Eq.~(\ref{eq-A}) is immediately 
transformed into an equation of motion of ${\bf E}(k)$ which
further provides the dispersion equation (\ref{dispersion-g}).

The dynamical information is contained in the polarization tensor 
$\Pi^{\mu \nu}(k)$ or, equivalently, in the permittivity tensor 
$\epsilon^{lm}(k)$ which can be derived either within the transport 
theory or diagrammatically \cite{Mrowczynski:2000ed}. The result is
\be
\label{epsilon}
\epsilon^{nm} (\omega, {\bf k}) = \delta^{nm} + 
{g^2 \over 2\omega} \int {d^3 p \over (2\pi )^3}
{ v^n \over \omega - {\bf k v} + i0^+} 
{\partial f({\bf p}) \over \partial p^l} 
\Bigg[ \Big( 1 - {{\bf k v} \over \omega} \Big) \delta^{lm}
+ {k^l v^m \over \omega} \Bigg] \,.
\ee
Since $\Pi^{\mu \nu}(k)$ and $\epsilon^{lm}(k)$ are unit matrices
in the colour space, the colour indices are suppressed here.

Substituting the permittivity (\ref{epsilon}) into 
Eq.~(\ref{dispersion-g}), one fully specifies the dispersion
equation (\ref{dispersion-g}) which provides a spectrum of
quasi-particle bosonic excitations. A solution $\omega ({\bf k})$ 
of Eq.~(\ref{dispersion-g}) is called {\it stable} when ${\rm Im}\omega \le 0$ 
and {\it unstable} when ${\rm Im}\omega > 0$. In the first case the amplitude 
is constant or it exponentially decreases in time while in the second 
one there is an exponential growth of the amplitude. In practice it 
appears difficult to find solutions of Eq.~(\ref{dispersion-g}) because 
of rather complicated structure of the tensor (\ref{epsilon}). However, 
the problem simplifies as we are interested in specific modes which are 
expected to be unstable. Namely, we look for solutions corresponding 
to the fluctuating current in the direction of the momentum surplus 
and the wave vector perpendicular to it. 

As previously, the momentum distribution is assumed to be elongated
in the $z$ direction, and consequently the fluctuating current also 
flows in this direction. The magnetic field has a non-vanishing component
along the $y$ direction and the electric filed in the $z$ direction. 
Finally, the wave vector is parallel to the axis $x$, see Fig.~1. We also 
assume that the momentum distribution obeys the mirror symmetry 
$f(-{\bf p}) = f({\bf p})$, and then the permittivity tensor has
only non-vanishing diagonal components. Taking into account all these 
conditions, one simplifies the dispersion equation (\ref{dispersion-g})
to the form
\be
\label{H-eq}
H(\omega) \equiv k^2_x - \omega^2 \epsilon^{zz}(\omega, k_x) = 0 \;,
\ee
where only one diagonal component of the dielectric tensor enters.

It appears that an existence of unstable solutions of Eq.~(\ref{H-eq}) 
can be proved without solving it. The so-called Penrose criterion 
\cite{Kra73}, which follows from analytic properties of the 
permittivity as a function of $\omega$, states that {\em the 
dispersion equation $H(\omega ) = 0$ has unstable solutions if} 
$H(\omega = 0) < 0$. The Penrose criterion was applied to the
equation (\ref{H-eq}) in \cite{Mrowczynski:xv} but a much more 
general discussion of the instability condition is presented in 
\cite{Arnold:2003rq}. Not entering into details, there exist
unstable modes if the momentum distribution averaged (with
a proper weight) over momentum length is anisotropic.

To solve the dispersion equation (\ref{H-eq}), the parton 
momentum distribution has to be specified. Several analytic
(usually approximate) solution of the dispersion equation with 
various momentum distributions can be found in 
\cite{Mrowczynski:xv,Romatschke:2003ms,Arnold:2003rq}.
An example of the numerical solution, which gives the unstable 
mode frequency in the full range of wave vectors is shown Fig.~2 
taken from \cite{Randrup:2003cw}. The momentum distribution is 
of the form
$$
f({\bf p}) \sim 
\frac{1}{(p_T^2 + \sigma_\perp^2)^3}\, 
{\rm e}^{-{p_z^2 \over 2\sigma_\parallel^2}}\;,
$$
where $p_\perp \equiv \sqrt{p_x^2 + p_y^2}$. The mode is pure imaginary
and $\gamma_k \equiv {\rm Im}\omega (k_\perp)$. The value of the
coupling is $\alpha_s \equiv g^2/4\pi =0.3$, 
$\sigma_\perp = 0.3\; {\rm GeV}$ and the effective parton 
density is chosen to be $6 \; {\rm fm}^{-3}$. As seen, there is 
a finite interval of wave vectors for which the unstable modes exist.

\begin{figure}[h]
\begin{minipage}{18pc}
\includegraphics[width=18pc]{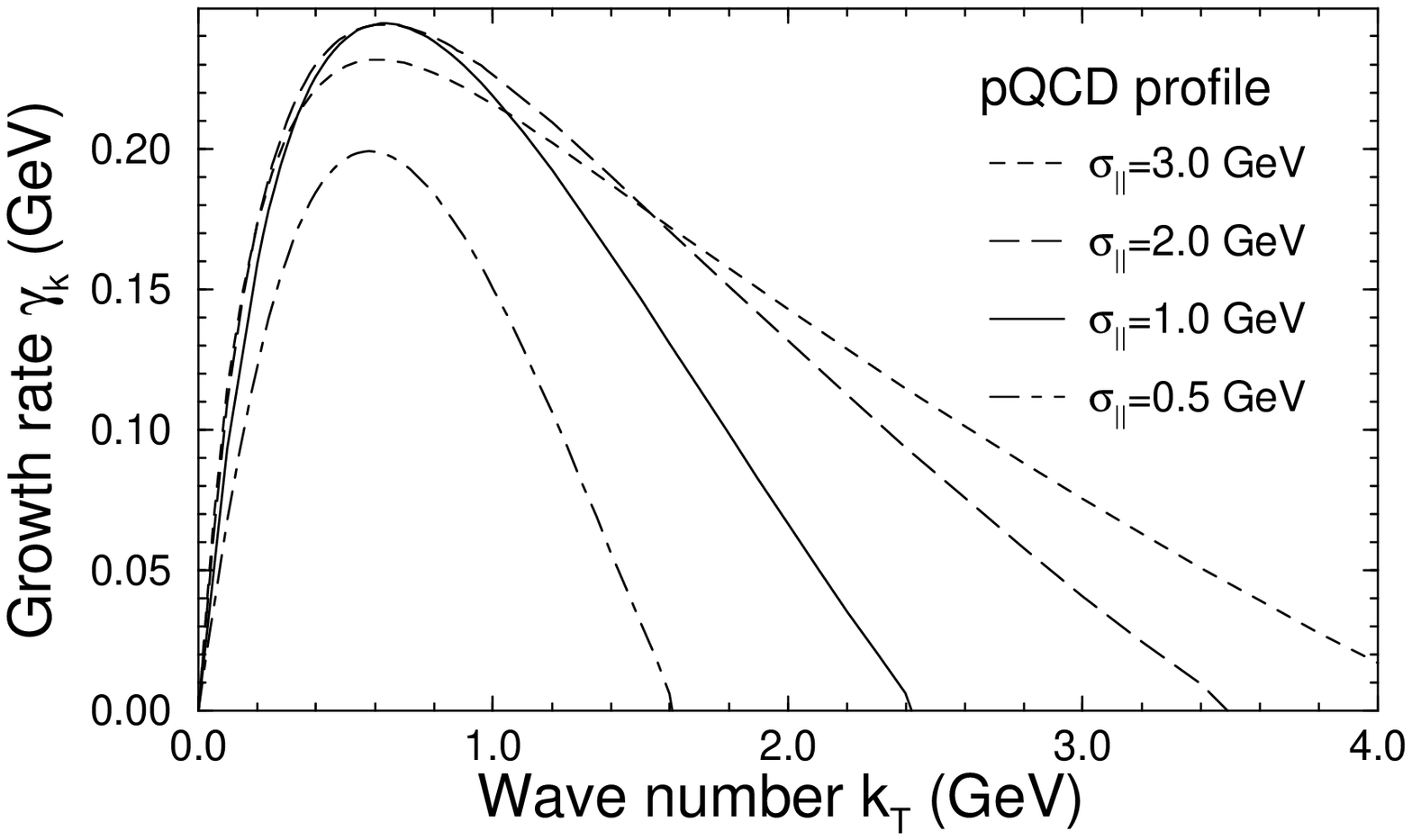}
\vspace{-0.7cm}
\caption{The growth rate of the unstable mode as a function of the
wave vector ${\rm k}=(k_\perp,0,0)$ for $\sigma_\perp=0.3~{\rm GeV}$
and 4 values of the parameter $\sigma_\parallel$ which controls system's 
anisotropy. The figure is taken from \cite{Randrup:2003cw}.}
\end{minipage}\hspace{2pc}%
\begin{minipage}{16pc}
\includegraphics[width=16pc]{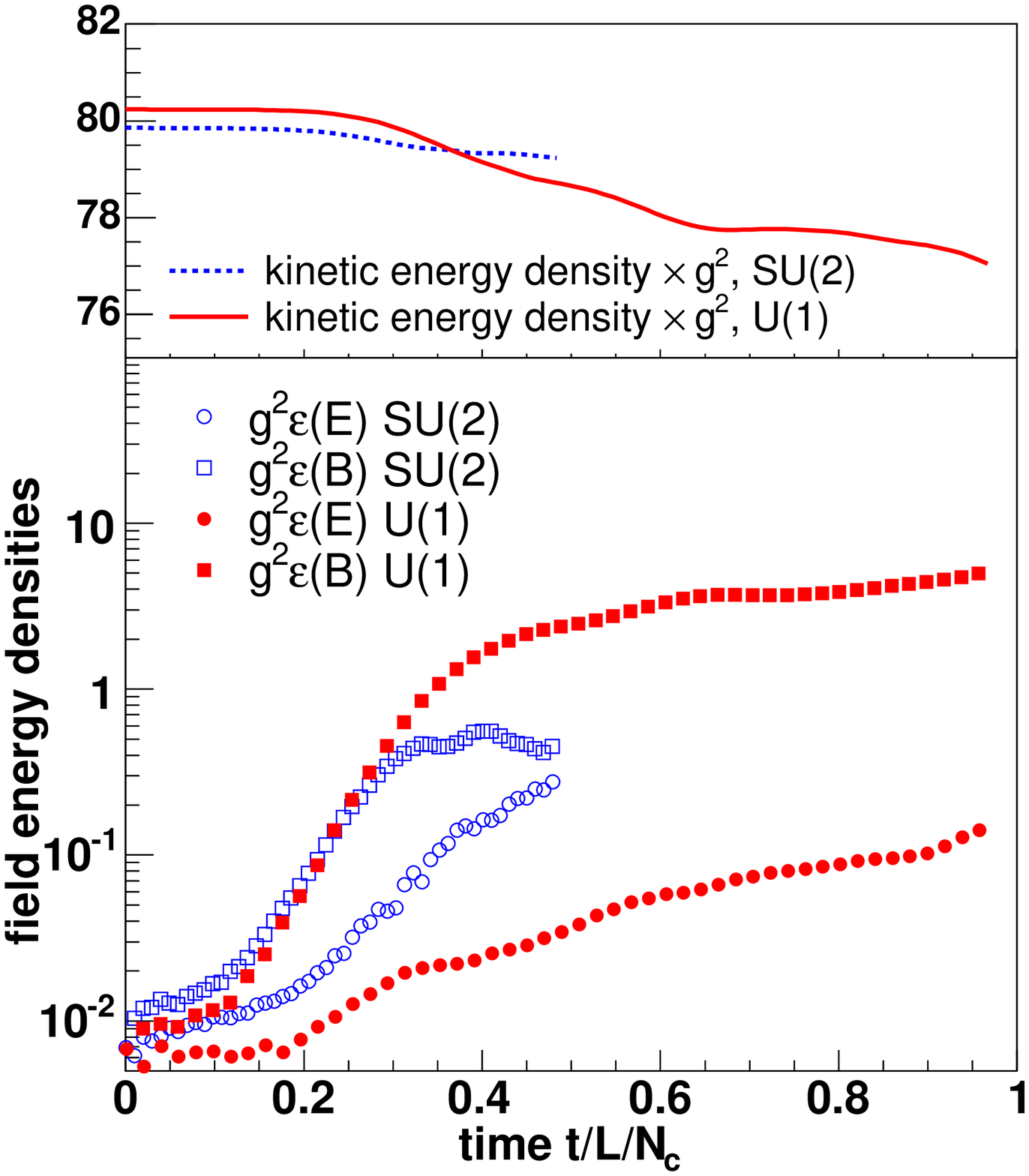}
\vspace{-0.7cm}
\caption{Time evolution of the kinetic energy of particles (upper panel)
and the energy stored in electric and magnetic fields (lower panel)
in ${\rm GeV}/{\rm fm}^3$ for the U(1) and SU(2) gauge groups. The figure
is taken from \cite{Dumitru:2005gp}.}
\end{minipage}
\end{figure}

%%%%%%%%%%%%%%%%%%%%%%%%%%%%%%%%%%%%%%%%%%%%%%%%%%%%%%%%%%%%%%%%%%%%%%%%%%%
                                                                                
\subsection{Growth of instabilities and abelianization}
                                                                                
%%%%%%%%%%%%%%%%%%%%%%%%%%%%%%%%%%%%%%%%%%%%%%%%%%%%%%%%%%%%%%%%%%%%%%%%%%%

A time evolution of a classical many-parton system interacting via
classical chromodynamic field has been studied in \cite{Dumitru:2005gp}.
Numerical simulations have been performed effectively in $1+1$ dimensions 
as the chromodynamic potentials depend on $t$ and $x$. The initial
field amplitudes are assumed to obey Gaussian white noise and the 
initial parton momentum distribution is
\be
\label{initial}
f({\bf p}) \sim
\delta(p_z) \;
{\rm e}^{-\frac{\sqrt{p_x^2 + p_y^2}}{p_{\rm hard}}}\;,
\ee
with $p_{\rm hard}= 10 \; {\rm GeV}$.

Fig.~3, which is taken from \cite{Dumitru:2005gp}, shows results of
the simulation corresponding to a lattice of physical size 
$L=40 \; {\rm fm}$. As seen, the amount of energy of the fields, 
which is initially much smaller than the kinetic energy of all 
particles, grows exponentially and the magnetic contribution dominates.
The simulation \cite{Dumitru:2005gp} indirectly confirms existence 
of the unstable magnetic modes in the system.  

Unstable modes cannot grow to infinity and the question arises 
what is a mechanism responsible for stopping the instability growth.
One suspects that non-Abelian non-linearities can play an important
role here. An elegant qualitative argument \cite{Arnold:2004ih} suggests 
that the non-linearities do not stabilize the unstable modes because 
the system spontaneously chooses an Abelian configuration in the course 
of the instability development. Let me explain the idea.

In the Coulomb gauge the effective potential of the unstable 
configuration has the form
\ban
V_{\rm eff}[{\bf A}^a] = - \mu^2 {\bf A}^a \cdot {\bf A}^a
+  \frac{1}{4} g^2 f^{abc} f^{ade}
({\bf A}^b {\bf A}^d) ({\bf A}^c {\bf A}^e) \;,
\ean
which is shown in Fig.~4 taken from \cite{Arnold:2004ih}. The first 
term (with $ \mu^2 >0$) is responsible for a very existence of the
instability. The second term, which comes from the Yang-Mills
lagrangian, is of pure non-Abelian nature. The term appears to be positive
and thus it counteracts the instability growth. However, the non-Abelian
term vanishes when the potential ${\bf A}^a$ is effectively Abelian, and
consequently, such a configuration corresponds to the steepest decrease 
of the effective potential. Thus, the system spontaneously abelianizes 
in the course of instability growth.

\begin{figure}
\begin{center}
\includegraphics[width=13cm]{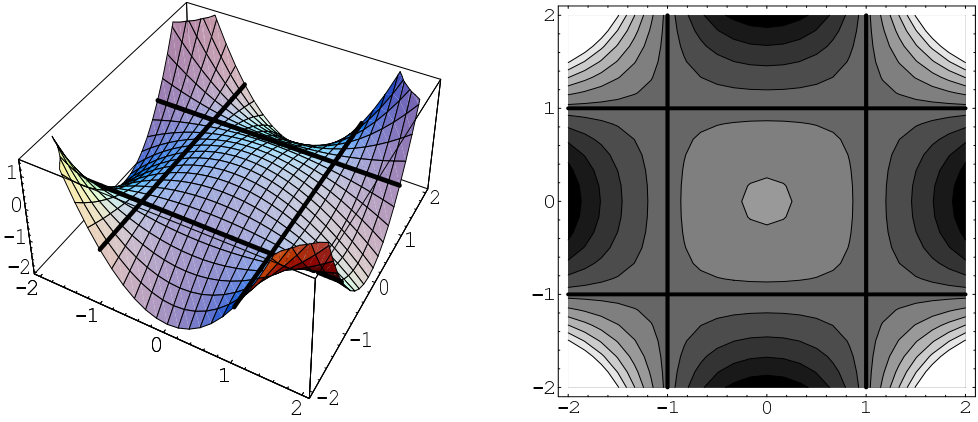}
\end{center}
\vspace{-0.5cm}
\caption{The effective potential of the unstable magnetic mode as
a function of two colour components of  ${\bf A}^a$ belonging to 
the SU(2) gauge group. The figure is taken from \cite{Arnold:2004ih}.}
\end{figure}

The effect of abelianization has been indeed found in numerical
simulations performed in the $1+1$ dimensions 
\cite{Rebhan:2004ur,Dumitru:2005gp,Arnold:2004ih}. As an example,
I show in Fig.~5 the result of fully classical simulation 
\cite{Dumitru:2005gp}. One observes in Fig.~5 taken from 
\cite{Dumitru:2005gp}, where 
\ban
\phi_{\rm rms} \equiv \sqrt{\int_0^L \frac{dx}{L} 
(A^a_yA^a_y + A^a_zA^a_z)}
\;, \;\;\;\;\;\;
\bar C \equiv \int_0^L \frac{dx}{L} 
\frac{\sqrt{ {\rm Tr}[(i[A_y,A_z])^2]}}{{\rm Tr}[A_y^2+A_z^2]} \;,
\ean
that the field commutator measured by $\bar C$ decreases in time, 
in spite of the field growth quantified by $\phi_{\rm rms}$. 

\begin{figure}[h]
\begin{minipage}{17.5pc}
\includegraphics[width=17.5pc]{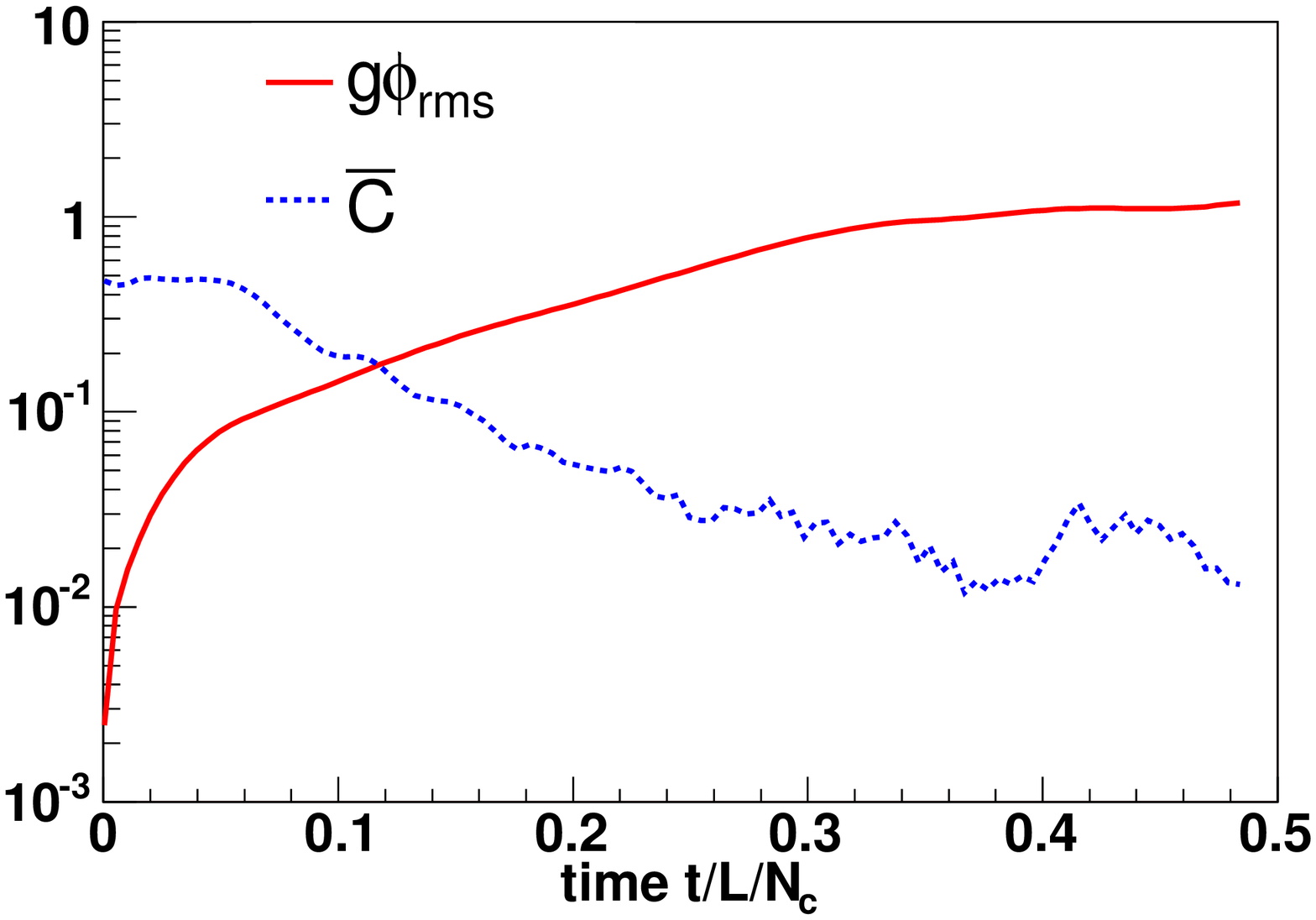}
\vspace{-0.8cm}
\caption{Temporal evolution of  $\bar C$ and $\phi_{\rm rms}$ measured
in GeV. The figure is taken from \cite{Dumitru:2005gp}.}
\end{minipage}\hspace{2pc}%
\begin{minipage}{18.5pc}
\vspace{-0.7cm}
\includegraphics[width=18.5pc]{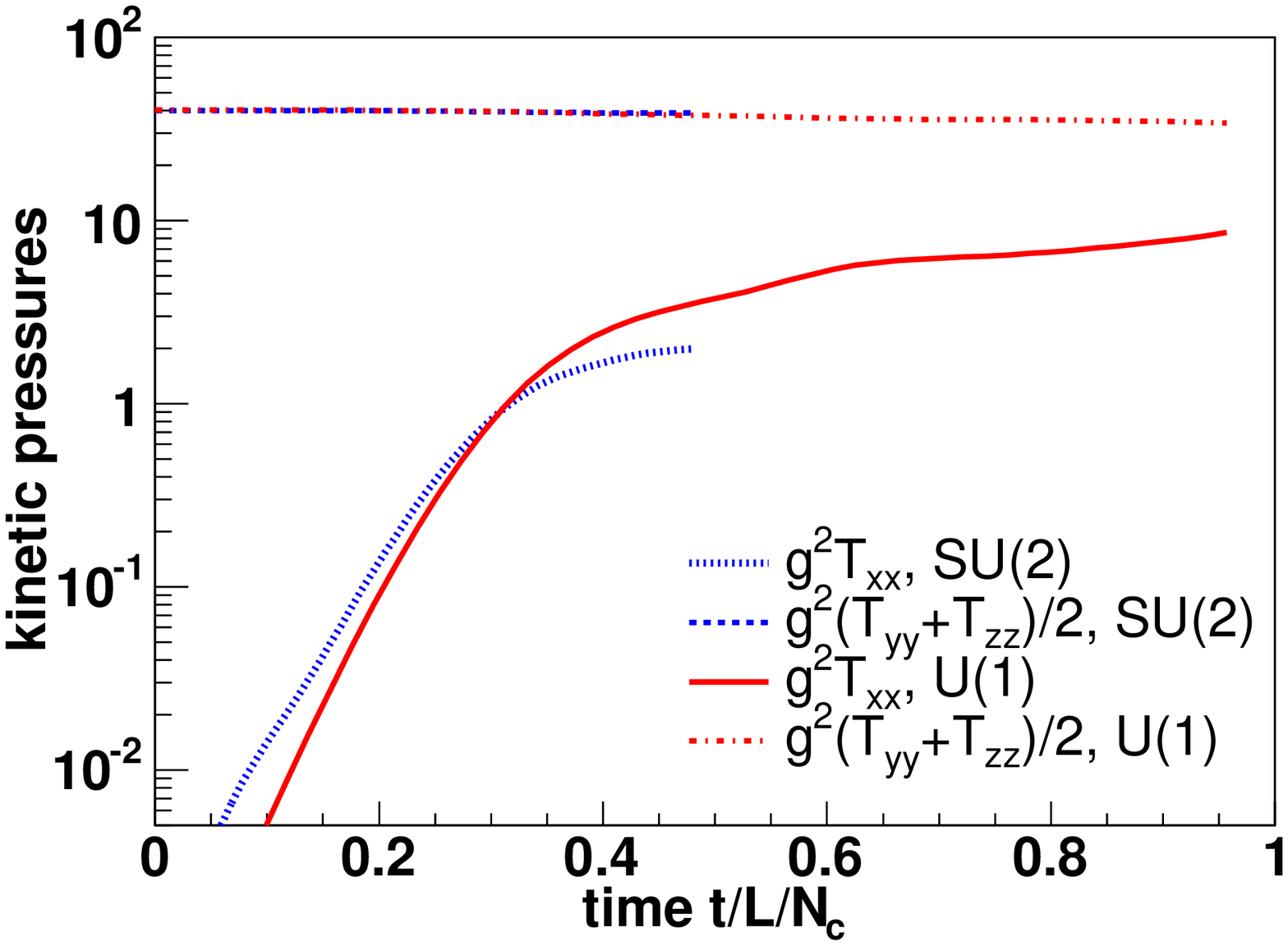}
\vspace{-0.8cm}
\caption{Temporal evolution of $T^{xx}$ and $(T^{yy}+T^{zz})/2$.
The figure is taken from \cite{Dumitru:2005gp}.}
\end{minipage}
\end{figure}

Very recent simulations performed in the $1+3$ dimensions \cite{Arnold:2005vb,Rebhan:2005re}, which utilise a complete
Hard Loop action for anisotropic systems \cite{Mrowczynski:2004kv}, 
show that the growth of unstable modes, which is initially 
exponential, becomes only linear at the later times. And
the abelianization works only  in the exponential period 
of instability development. However, it might well be that
the abelianization becomes more efficient when the dynamical
effects beyond the Hard Loop approximation are taken into
account \cite{Manuel:2005mx}.

%%%%%%%%%%%%%%%%%%%%%%%%%%%%%%%%%%%%%%%%%%%%%%%%%%%%%%%%%%%%%%%%%%%%%%%%%%%
                                                                                
\subsection{Isotropization}
                                                                                
%%%%%%%%%%%%%%%%%%%%%%%%%%%%%%%%%%%%%%%%%%%%%%%%%%%%%%%%%%%%%%%%%%%%%%%%%%%

When instabilites grow the systems becomes more isotropic 
because the Lorentz force acts on particle's momenta and 
the growing fields generate an extra momentum. 

To explain the mechanism I assume, as previously, that initially 
there is a momentum surplus in the $z$ direction. The fluctuating 
current tends to flow in the $z$ direction with the wave vector pointing 
in the $x$ direction. Since the magnetic field has a $y$ component, 
the Lorentz force, which acts on partons flying along the $z$ axis, 
pushes the partons in the $x$ direction where there is a momentum 
deficit.

The effect of isotropization due to the action of the Lorentz force
is nicely seen in the classical simulation \cite{Dumitru:2005gp}.
In Fig.~6, which is taken from \cite{Dumitru:2005gp}, there are shown
diagonal components of the energy-momentum tensor
$$
T^{\mu \nu} = \int \frac{d^3 p}{(2\pi)^3}\frac{p^\mu p^\nu}{E_p}
f({\bf p}) \;.
$$
The initial momentum distribution is given by Eq.~(\ref{initial}),
and consequently $T^{xx}=0$ at $t=0$. As seen in Fig.~6, $T^{xx}$
exponentially grows.
 
The system isotropizes not only due to the effect of the Lorentz
force but also due to the momentum carried by the growing field. 
As explained in detail in \cite{Mrowczynski:xv}, the momentum of 
the field is oriented along the wave vector which points in the 
direction of the momentum deficit. This effect has not been 
numerically studied yet but it is clear that the effect is comparable 
to that of Lorentz force only for suffuciently large field amplitudes.

%%%%%%%%%%%%%%%%%%%%%%%%%%%%%%%%%%%%%%%%%%%%%%%%%%%%%%%%%%%%%%%%%%%%%%%%%%%
                                                                                
\section{Azimuthal fluctuations}
                                                                                
%%%%%%%%%%%%%%%%%%%%%%%%%%%%%%%%%%%%%%%%%%%%%%%%%%%%%%%%%%%%%%%%%%%%%%%%%%%

In the first part of my talk I have argued that the quark-gluon plasma 
becomes isotropic fast due to the magnetic instabilities. And it has 
been recently observed \cite{Arnold:2004ti} that the system with isotropic 
momentum distribution manifests a hydrodynamic collective behaviour.
The question arises whether such an approximate hydrodynamics can be
distinguished from the real hydrodynamics describing a system which is 
in a local thermodynamic equilibrium. In the second part of my talk 
I propose to address the question by studying the azimuthal fluctuations. 

In relativistic heavy-ion collisions both at CERN SPS and BNL RHIC,
one observes a sizable elliptic flow which is quantified by the second 
angular harmonics $v_2$ of the azimuthal distribution of final state hadrons 
\cite{Alt:2003ab,Adler:2002pu,Adler:2003kt,Back:2004zg}. The phenomenon, 
which is sensitive to the collision early stage \cite{Sorge:1998mk} when 
the interaction zone is of the almond shape, is naturally explained within 
a hydrodynamics as a result of large density gradients 
\cite{Ollitrault:bk,Kolb:2000sd,Teaney:2001av,Hirano:2001eu}.
Hydrodynamic description requires that the system under study
is in a local thermodynamical equilibrium. However, an approximate 
hydrodynamic behaviour occurs, as argued in \cite{Arnold:2004ti},
when the momentum distribution of liquid components is merely 
isotropic in the local rest frame. The point is that the structure 
of the ideal fluid energy-momentum tensor {\it i.e.} 
$T^{\mu \nu} = (\varepsilon + p) \, u^{\mu} u^{\nu} -p \, g^{\mu \nu}$, 
where $\varepsilon$, $p$ and $u^{\mu}$ is the energy density, 
pressure and hydrodynamic velocity, respectively, holds for an
arbitrary, though isotropic momentum distribution. $\varepsilon$ 
and $p$ are then not the energy density and pressure but the 
moments of the distribution function which are equal the energy 
density and pressure in the equilibrium limit. Since the tensor 
$T^{\mu \nu}$ obeys the continuity equation 
$\partial_\mu T^{\mu \nu} =0$, one gets an analogue of the Euler
equation. However, due to the lack of thermodynamic equilibrium
there is no entropy conservation and the equation of state is
missing.

Usually, non-equilibrium fluctuations are significantly smaller
than the equilibrium fluctuations of the same quantity. A specific
example of such an situation has been discussed in Sec.~2.1. 
Therefore, I expect that the fluctuations of $v_2$ produced in 
the course of real hydrodynamic evolution are significantly smaller 
than those generated in the non-equilibrium quark-gluon plasma 
which is merely isotropic. It should be stressed here that the 
elliptic flow is generated in the collision early stage. Thus, 
I propose to carefully measure the fluctuations of $v_2$ as 
discussed in \cite{Mrowczynski:2002bw}. Since such a measurement 
is rather difficult, I also consider an integral measurement 
of azimuthal fluctuations proposed in \cite{Mrowczynski:1999vi} 
which can also help to distinguish the equilibrium from 
non-equilibrium fluctuations.

%%%%%%%%%%%%%%%%%%%%%%%%%%%%%%%%%%%%%%%%%%%%%%%%%%%%%%%%%%%%%%%%%%%%%%%%%%%
                                                                                
\subsection{Elliptic flow fluctuations}
                                                                                
%%%%%%%%%%%%%%%%%%%%%%%%%%%%%%%%%%%%%%%%%%%%%%%%%%%%%%%%%%%%%%%%%%%%%%%%%%%

In my discussion of $v_2$ fluctuations I follow \cite{Mrowczynski:2002bw} 
where the standard method \cite{Voloshin:1994mz,Poskanzer:1998yz} to 
measure the elliptic flow was used. The method focuses on the 
angular distributions relative to direction of the impact parameter. 
The experimental procedure splits in two steps which should be as 
independent as possible. In the first step, one determines the impact 
parameter direction $\psi_R$, while in the second step, one constructs 
the distribution of azimuthal angle relative to $\psi_R$ and one computes 
the Fourier coefficients. 

The one-particle distribution in a single event can be written as
\be 
\label{Fourier}
P_{\rm ev}(\phi) = {1 \over 2\pi} \;
\Big[1 + 2\sum_{n=1}^{\infty} 
v_n {\rm cos}\big(n(\phi- \psi_R)\big) \Big] 
\Theta(\phi) \, \Theta(2\pi - \phi) \;.
\ee
Since the reaction plane is never reconstructed precisely and the 
real reaction plane angle $\psi_R$ deviates from the estimated angle 
$\psi_E$, the $n-$th Fourier amplitude $v_n$ is determined as
$$
v_n = {1 \over R_n} \; \overline{{\rm cos}\big( n(\phi -\psi_E)\big)} \;,
$$
where $R_n \equiv {\rm cos}\big( n(\psi_R -\psi_E)\big)$ is the 
reaction plane resolution factor and $\overline{\cdots}$ denotes
averaging over particles from a single event.

Let me now consider an ensemble of events with every event representing
a single nucleus-nucleus collision at fixed value (not direction) of
the impact parameter. The angle $\psi_R$ (and $\psi_E$) obviously varies 
form event to event. The question is how to detect the event-by-event 
fluctuations of the Fourier amplitudes $v_n$. According to 
\cite{Poskanzer:1998yz}, the amplitude averaged over events is 
defined as 
$$
\langle v_n \rangle \buildrel \rm def \over = 
{\Big\langle  \overline{{\rm cos}\big( n(\phi -\psi_E)\big)} \Big\rangle 
\over \langle R_n \rangle } \;,
$$
where $\langle \cdots \rangle$ denotes averaging over events.
We define the second moment as 
$$
\langle v_n^2 \rangle \buildrel \rm def \over = 
{ 1 \over \langle R_n \rangle^2 } 
\Big\langle \overline{{\rm cos}\big( n(\phi -\psi_E)\big)}^2 
\Big\rangle \;,
$$
and the fluctuations are
$$
{\sf Var}(v_n) \equiv \langle v_n^2 \rangle - \langle v_n \rangle^2 
= { 1 \over \langle R_n \rangle^2 } 
\Big(\Big\langle \overline{{\rm cos}\big( n(\phi -\psi_E)\big)}^2 
\Big\rangle 
- \Big\langle \overline{{\rm cos}\big( n(\phi -\psi_E)\big)}
\Big\rangle^2 \Big) \;.
$$

There are several sources of trivial $v_2$ fluctuations which
are not related to the system's dynamics of interest. I start 
with the fluctuations caused by a varying number of particles
used to determine $\psi_E$ and $v_2$. I assume here that the 
Fourier amplitudes $v_n$ do {\em not} change from event to event
and that the only correlations in the system are those due to 
the flow. Then, the azimuthal distribution of $N$ particles is 
a product of $N$ single particle distributions
\be
\label{noise-distr}
P^N_{\rm ev}(\phi_1,\phi_2, \cdots, \phi_N) = {\cal P}_N
P_{\rm ev}(\phi_1) \: P_{\rm ev}(\phi_2) \; \cdots \; 
P_{\rm ev}(\phi_N) \;,
\ee
where ${\cal P}_N$ is the multiplicity distribution while all 
distributions $P_{\rm ev}(\phi_i)$ are given by Eq.~(\ref{Fourier}).
The single particle distributions $P_{\rm ev}(\phi_i)$ are correlated 
to each other because of the common angle $\psi_R$. 

Using the distribution (\ref{noise-distr}), one finds (neglecting
$v_4$) the variance of $v_2$ as
\be 
\label{noise2}
{\sf Var}(v_2) = {1\over 2 \langle R_2 \rangle^2 \langle N \rangle} 
+ \langle v_2 \rangle^2 \;
{\langle R_2^2 \rangle - \langle R_2 \rangle^2 \over 
\langle R_2 \rangle^2} \;,
\ee
which holds for $\langle N \rangle \gg 1$ and small multiplicity 
fluctuations. The second term in r.h.s of Eq.~(\ref{noise2}) 
appears to be much smaller than the first one as
$\langle R_2^2 \rangle - \langle R_2 \rangle^2 \sim 
\langle M \rangle^{-2}$, where $M$ is the number of particles
used to determine the reaction plane and $M$ is assumed to be of 
the same order as $N$. Thus, the statistical noise contribution
to $\delta v_2 \equiv \sqrt{{\sf Var}(v_2)}$ is finally estimated 
as
\be 
\label{noise1}
\delta v_2 = {1\over \langle R_2 \rangle \sqrt{2 \langle N \rangle }}\;.
\ee

As well known, $\langle v_2 \rangle$ strongly depends on the collision
impact parameter $b$. Using the parameterisation of this dependence
given in \cite{Adler:2002pu}, one computes $\delta v_2$ as  
$(d \langle v_2 \rangle/db)\delta b$. The impact parameter is 
measurable through the multiplicity $N_p$ of participating nucleons. 
$N_p$ is directly related to $b$. For $b \approx 10$ fm, when the 
flow in Au-Au collisions is maximal, the $v_2$ fluctuations due to 
the impact parameter variation vanish because 
$d \langle v_2 \rangle /db = 0$. For $b = 5$ fm, 
where $\langle v_2 \rangle \approx 0.03$, 
one finds $\delta v_2 \approx 8 \cdot 10^{-4}\: \delta N_p$. 
When $\delta N_p = 30$ and $\langle N \rangle = 500$, the magnitude
of the $v_2$ fluctuations caused by the impact parameter variation 
is approximately equal to that of the statistical noise. 

The next source of trivial $v_2$ fluctuations is a variation 
of thermodynamic parameters. The contribution caused by 
the particle number fluctuations can be estimated as
$$
\delta v_2 = 
{d \langle v_2 \rangle \over d\langle N \rangle }\; \delta N 
=  \langle v_2 \rangle \; {\delta N \over \langle N \rangle } P \;,
$$
where the effective power $P$ is
$$
P \equiv 
{d {\rm ln} \langle v_2 \rangle \over d {\rm ln} \langle N \rangle} 
= {\langle N \rangle \over \langle v_2 \rangle} \;
{d \langle v_2 \rangle \over d \langle N \rangle} \;.
$$
Assuming the poissonian character of multiplicity fluctuations,
one obtains
\be 
\label{hydro-2}
\delta v_2 = {\langle v_2 \rangle \over \sqrt{\langle N \rangle}} \; P \;.
\ee
The value of the index $P$ can be estimated within the hydrodynamic
model \cite{Teaney:2001av} as $P \approx 0.4$. Comparing 
Eqs.~(\ref{noise1}) and (\ref{hydro-2}), one finds that the ratio 
of the thermodynamic fluctuations to the statistical noise, is 0.04 
for $\langle v_2 \rangle = 0.07$ and $P = 0.4$. Thus, the thermodynamic 
fluctuations are much smaller than the statistical noise. 

Concluding this section, I propose to perform a systematic measurement
of event-by-event $v_2$ fluctuations for the centrality corresponding
to the maximal elliptic flow. Then, the fluctuations caused by the
impact parameter variation vanish or at least they are very small. 
If the flow is built up in the course of hydrodynamic evolution of 
the equilibrium system, $v_2$ should be dominated by the statistical 
noise related to the finite particle number. The noise can be identified 
due to the characteristic $1/\langle N \rangle$ dependence.

%%%%%%%%%%%%%%%%%%%%%%%%%%%%%%%%%%%%%%%%%%%%%%%%%%%%%%%%%%%%%%%%%%%%%%%%%%%
                                                                                
\subsection{$\Phi$ measure of azimuthal fluctuations}
                                                                                
%%%%%%%%%%%%%%%%%%%%%%%%%%%%%%%%%%%%%%%%%%%%%%%%%%%%%%%%%%%%%%%%%%%%%%%%%%%

Since a measurement of $v_2$ fluctuations discussed in the previous
section is rather difficult, I suggest to consider a much simpler
integral measurement of azimuthal fluctuations \cite{Mrowczynski:1999vi},
using the so-called $\Phi$ measure introduced in \cite{Gazdzicki:ri}.

The correlation (or fluctuation) measure $\Phi$ is defined as follows.
One defines the variable $z\buildrel \rm def \over = x - \overline{x}$,
where $x$ is a single particle characteristics such as the particle 
transverse momentum or the azimuthal angle. In this section the overline
does not denote averaging over particles from a single event but averaging 
over a single particle inclusive distribution. $x$ is identified here 
with the particle azimuthal angle. The event variable $Z$, which is 
a multiparticle analog of $z$, is defined as 
$Z \buildrel \rm def \over = \sum_{i=1}^{N}(x_i - \overline{x})$, 
where the summation runs over particles from a given event. By construction, 
$\langle Z \rangle = 0$. The measure $\Phi$ is finally defined as
$$
\Phi \buildrel \rm def \over = 
\sqrt{\langle Z^2 \rangle \over \langle N \rangle} 
- \sqrt{\overline{z^2}} \;.
$$
$\Phi$ obviously vanishes in the absence of any inter-particle correlations.
Other properties of $\Phi$ are discussed in \cite{Mrowczynski:1999un}.

The $\Phi$ measure is sensitive to the azimthal fluctuations caused 
by the transverse collective flow. Let me compute it, assuming
that the only correlations present in the system are due to the
collective flow. The inclusive $\phi$ distrubtion, which is flat in 
the range $[0, 2\pi]$, provides $\overline \phi = \pi$ and 
$\overline{\phi^2} = {4 \over 3}\pi^2$, and consequently, 
$\overline{z^2} = {1 \over 3}\pi^2$. Since 
$Z = \sum_{i=1}^{N}(\phi_i - \overline \phi)$, one computes
$\langle Z^2 \rangle$, using the event distribution (\ref{Fourier}),
as
\ba
\label{Z2}
\langle Z^2 \rangle = \int_0^{2\pi} {d\psi_R \over 2\pi} \sum_N {\cal P}_N
\int_0^{2\pi} d\phi_1 \dots \int_0^{2\pi} d\phi_N\; 
P_{\rm ev}(\phi_1)\; \dots \; P_{\rm ev}(\phi_N) \;
(\phi_1 + \dots + \phi_N - N\overline \phi )^2 \,.
\ea
The averaging over the amplitudes $v_n$, which is not shown here, is 
implied. At first glance, the multi-particle distribution from 
Eq.~(\ref{Z2}) might look as a simple product of one-particle 
distributions. One should note however that every $P_{\rm ev}(\phi)$ 
depends on the reaction plane angle $\psi_R$, and the integration 
over $\psi_R$ leads to the {\it correlated} multi-particle distribution.

After elementary calculation, one finds for small $v_n$, poissonian
multiplicity distribution and $\langle N \rangle \gg 1$, an approximate 
expression of interest
\be
\label{phi-col}
\Phi \cong {3 \over \pi^2} \, \langle N \rangle \, 
\langle \: \sum_{n=1}^{\infty} \big({v_n \over n} \big)^2 \rangle \;.
\ee
If all amplitudes $v_n$ except $v_2$ vanish, as it approximately 
happens in the central rapidity domain, and $v_2$ equals a unique 
value 0.07, Eq.~(\ref{phi-col}) for $\langle N \rangle = 500$ gives 
$\Phi=0.2$.

As already mentioned, the transverse flow is far not the only source
of the azimuthal fluctuations. In particular, the effect of quantum 
statistics contribute here. To estimate the effect one computes $\Phi$ 
in the equilibrium ideal quantum gas. The result reads 
\cite{Mrowczynski:1999vi}
\be
\label{phi-phi}
\Phi = {\pi \over \sqrt{3}}
\bigg( \sqrt{{\widetilde \rho \over \rho}} - 1 \bigg)\;,
\ee
where
$$
\rho \equiv \int{d^3p \over (2\pi )^3} \;
{1 \over \lambda^{-1}e^{\beta E_p} \pm 1} \;,
\;\;\;\;\;\;\;
\widetilde \rho \equiv \int{d^3p \over (2\pi )^3} \;
{\lambda^{-1}e^{\beta E_p}
\over (\lambda^{-1}e^{\beta E_p} \pm 1)^2} \;,
$$
$\lambda$ denotes the fugacity, the upper sign is for fermions
and the lower one for bosons. As seen, $\Phi$ is independent of 
the system's volume and of the number of particle's internal degrees 
of freedom. For massless bosons with vanishing chemical potential, 
Eq.~(\ref{phi-phi}) gives $\Phi \approx 0.3$ for any $T$. More 
realistic calculations \cite{Mrowczynski:1999vi} provide 
$\Phi \approx 0.06$ for chemically equilibrated pions at $T = 150$ MeV. 
 
I conclude this section by saying that a measured value of $\Phi$, 
which would significantly exceed predictions of Eq.~(\ref{phi-col})
with non-fluctuating amplitudes $v_n$, would be an obvious signal 
of sizable dynamical fluctuations.

%%%%%%%%%%%%%%%%%%%%%%%%%%%%%%%%%%%%%%%%%%%%%%%%%%%%%%%%%%%%%%%%%%%%%%%%%%%
                                                                                
\section{Final remarks}
                                                                                
%%%%%%%%%%%%%%%%%%%%%%%%%%%%%%%%%%%%%%%%%%%%%%%%%%%%%%%%%%%%%%%%%%%%%%%%%%%

The magnetic instabilities provide a plausible mechanism responsible
for a surprisingly short equilibration time observed in relativistic 
heavy-ion collisions. Fast isotropization is a distinctive feature
of the mechanism. It is certainly desirable to look for experimentally 
detectable signals of the instabilities driven thermalization. In my 
talk I have proposed to study azimuthal fluctuations, in particular 
the event-by-event fluctuations of the elliptic flow which is generated
at the collision early stage. I have not been able to present a quantitative
prediction but observation of sizeable dynamical fluctuations would be 
a strong argument that behind a smooth hydrodynamic evolution there 
is a violent phenomenon of plasma instabilities. 

\vspace{0.5cm}

%************************************************************************|
%
%                            Bibliography
%
%************************************************************************|

\end{document}